\documentclass[a4paper,11pt]{article}
\pdfoutput=1 
\usepackage{graphicx}
\usepackage{jinstpub} 
\usepackage{lineno}

\title{\boldmath Follow-up of the IceCube alerts with the Baikal-GVD telescope}

\author[a,1]{V.Y.~Dik} 

\affiliation[a]{Joint Institute for Nuclear Research, Dubna, Russia}

\note{Corresponding author.}

\emailAdd{viktoriya@jinr.ru}

\collaboration[c]{on behalf of the Baikal-GVD collaboration$^*$\note[*]{A complete list of authors can be found at the end of the proceedings}}

\abstract{The high-energy muon neutrino events of the IceCube telescope, that are triggered as neutrino alerts in one of two probability ranks of astrophysical origin, "gold" and "bronze", have been followed up by the Baikal-GVD in a fast quasi-online mode since September 2020.  Search for correlations between alerts and GVD events reconstructed in two modes, muon-track and cascades (electromagnetic or hadronic showers), for the time windows $ \pm $ 1 h and $ \pm $ 12 h does not indicate statistically significant excess of the measured events over the expected number of background events.  Upper limits on the neutrino fluence will be presented for each alert. }


\arxivnumber{2107.14303}

\proceeding{Very Large Volume Neutrino Telescope 2021 conference\\
18-21 May, 2021\\
Valencia / online
}

\begin{document}
\maketitle
\flushbottom

\section{Introduction}
\label{sec:intro}
At present, Baikal Gigaton Volume Detector (GVD)~\cite{Baikal2020} is the largest operating neutrino telescope at the Northern Hemisphere (51.77$^{\circ}$N, 104.42$^{\circ}$E), being at the stage of deployment with a gradual increase of its effective volume to the scale of 1 km$ ^{3}$. Currently, the GVD effective volume is about of 0.4 km$^3$ for electron neutrinos detection in an energy range of 100 TeV. 
Participating in multi-messenger program~\cite{SuvICRC}, Baikal-GVD follows up the neutrino alerts of the ANTARES telescope since December 2018~\cite{ANT_GVD}. Since September 2020 it also follows up the neutrino alerts of the IceCube telescope~\cite{Astr_let} in a quasi-online regime using the Gamma-ray Coordinates Network (GCN) circular~\cite{GCN}. 
The search for correlation between alerts on the IceCube astrophysical neutrino candidates received between September 2020 and May 2021 and events reconstructed in Baikal-GVD data in muon and cascade modes is presented in this contribution.

\section{Baikal-GVD data transfer in quasi-online mode}
\label{sec:2}
The modular system of the GVD telescope comprises 8 clusters as of 2021. Each cluster consists of 288 optical modules (OMs) with photodetectors designed to detect the Cherenkov light emitted when showers and muons are generated by high-energy neutrino interactions in water. Each cluster is a functionaly separate detector with photon registration, triggering, calibration, positioning and data acquisition systems. 
The trigger conditions are met when the threshold values $Q_{high}$ ($\sim$ 3--5 p.e.) and $Q_{low}$ ($\sim$ 1--2 p.e.) of the input signal amplitudes are exceeded on two neighbour optical modules (with the distance $\pm$ 15 m) 
for a time window of 100 ns. 
If an event passes the trigger, data from corresponding OMs are transferred to the Data Acquisition system on Baikal lake shore. Raw data from the shore are transferred to the storage facility at the Joint Institute for Nuclear Research in Dubna that is at about 4300 km away from GVD, via radio channel and Internet without preprocessing. Data processing is carried out with Baikal Analysis and Reconstruction Software framework~\cite{BARS}. It includes a set of programs written in C++ and Python. A special particularity of the C++ package based on ROOT is that the output data of one program are used as the input data to another programs. The task of the Python package based on the Luigi package~\cite{luigi} is to launch the C++ programs in the correct sequence and to monitor their execution. Data processing with user driven analysis pass in quasi-online mode. When data are being transferred from the shore to the storage facility, a delay of less than a minute can be possible. When the data are being processed, the delay increases between 3 to 5 h depending on the amount of luminescence noise~\cite{Astr_let}.

\section{IceCube alerts follow-up with Baikal-GVD}
\label{sec:3}
From September 2020 to May 2021 Baikal-GVD has received twenty alerts from the IceCube telescope located at the South Pole through GCN circular \cite{GCN}. Each alert refers to a muon track event in two categories of probability to be astrophysical origin: either type "gold"\ with signalness to be not less than 50\% or type "bronze"\ with signalness of at least 30\%. Angular accuracy of the muon reconstruction is about 0.5$^{\circ}$ in "gold"\ event and up to 2.5$^{\circ}$ in "bronze"\ one~\cite{IC_Gold_Bronze}. The angular resolution in one Baikal-GVD single cluster for muon (cascade) reconstruction has a median value of 1.3$^\circ$ (4.5$^\circ$). The median values have been accounting in search for follow-up events within a half open cone towards each alert direction. In fast regime reconstruction
we have used the cone of 5$^{\circ}$ both for shower and track-like events, so it covers contribution from IceCube track resolutions.

\begin{figure}[htbp]
	\centering 
	\includegraphics[width=.39\textwidth]{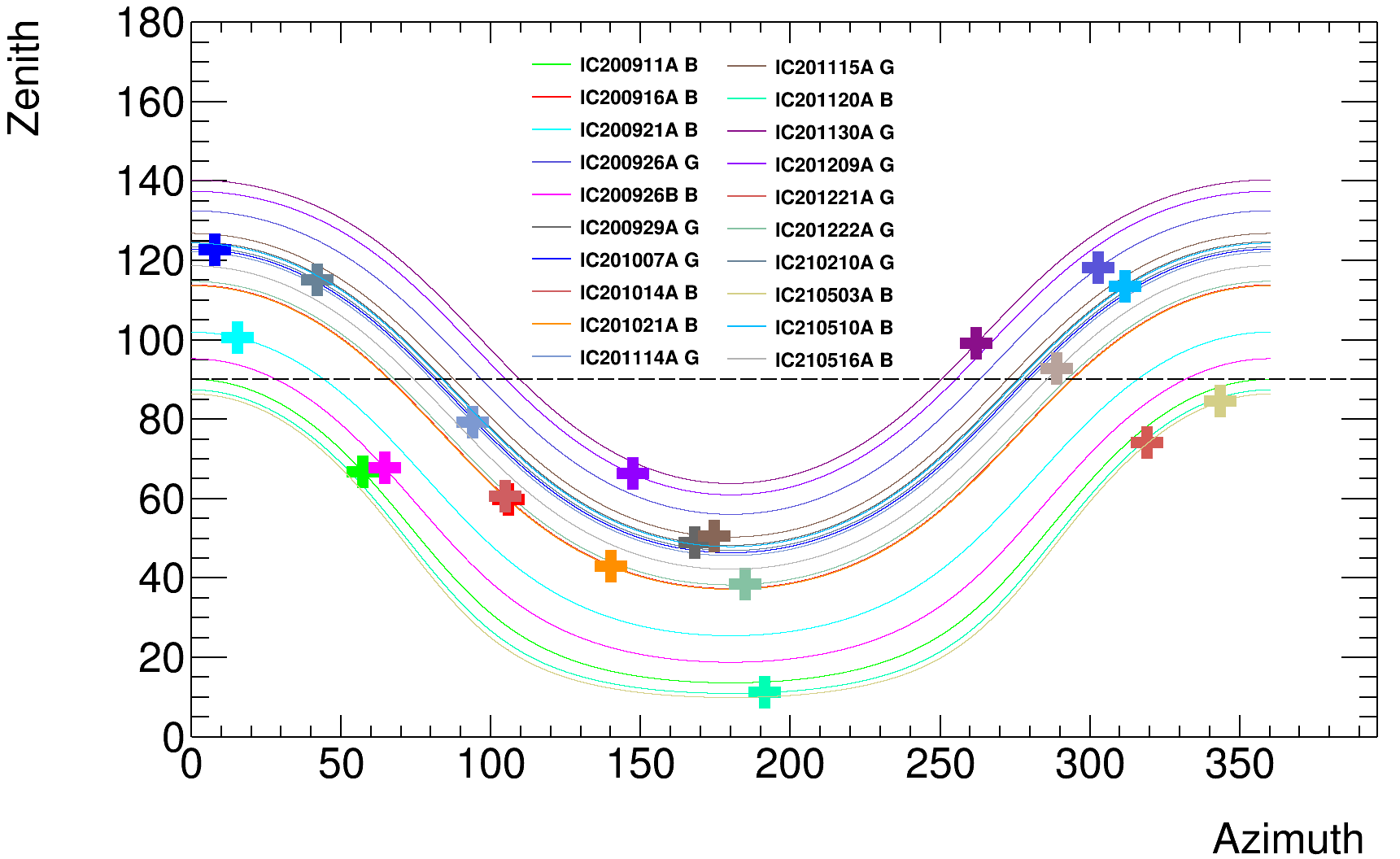}
    \includegraphics[width=.35\textwidth]{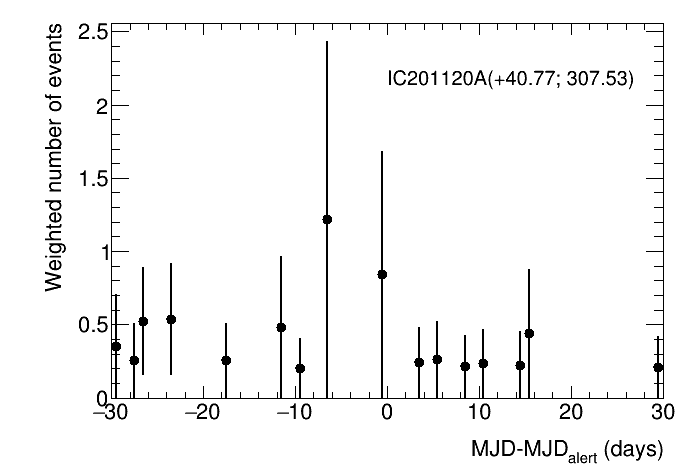}
	\caption{\label{fig:20_al} Left: Diurnal trajectories of the fixed equatorial coordinates of neutrino alerts in Baikal-GVD horizontal coordinates. The swiss cross signs show start points of sources, when they were observed by IceCube. The symbols "G" and "B" reffer to "gold" and "bronze" ranks of the IC alerts. The dotted line separates sources locations above horizon and below horizon. Right: Distribution of the reconstructed Baikal-GVD cascade events passing the selection criteria with $ \psi < 5^{\circ}$ from IC201120A for a time window of $\pm$ 30 days around the alert time. The events are weighted as w=1/$\psi$. } 
\end{figure}

Figure \ref{fig:20_al} (left) presents the visibility of followed up IceCube track events in interval $\pm$ 12 h from the alert time in horizontal coordinates for Baikal-GVD site. Seven events have a zenith angle between 92$^\circ$ and 124$^\circ$, while the others correspond to downward going events in Baikal-GVD. The track reconstruction of events near the horizon has a low quality in the single cluster mode~\cite{Safronov_ICRC2021} considering the current quality criteria of neutrino events in Baikal-GVD~\cite{Zaborov_ICRC2021}. The criteria for near-horizon muon track-like events are now at the research stage for both single cluster and multi cluster events. As preliminary result, no common events were identified in reconstructed track events within time windows of multi cluster coincidences.

A search for correlation in time and in celestial coordinates between IceCube alerts and Baikal-GVD events reconstructed in cascade mode have been done. We selected the shower-like events that satisfied a set of quality criteria \cite{casc_area1}, \cite{casc_area2} and had a number of triggered OMs larger than 7. Example of search for neutrino around IC201120A is displayed in Figure \ref{fig:20_al} (right), where considered distribution of GVD cascade events is presented for time interval $ \pm$ 30 days. For final results we considered time windows $\pm$ 1 h, $\pm$ 12 h. 
\begin{figure}[htbp]
	\centering 
	\includegraphics[width=.326\textwidth]{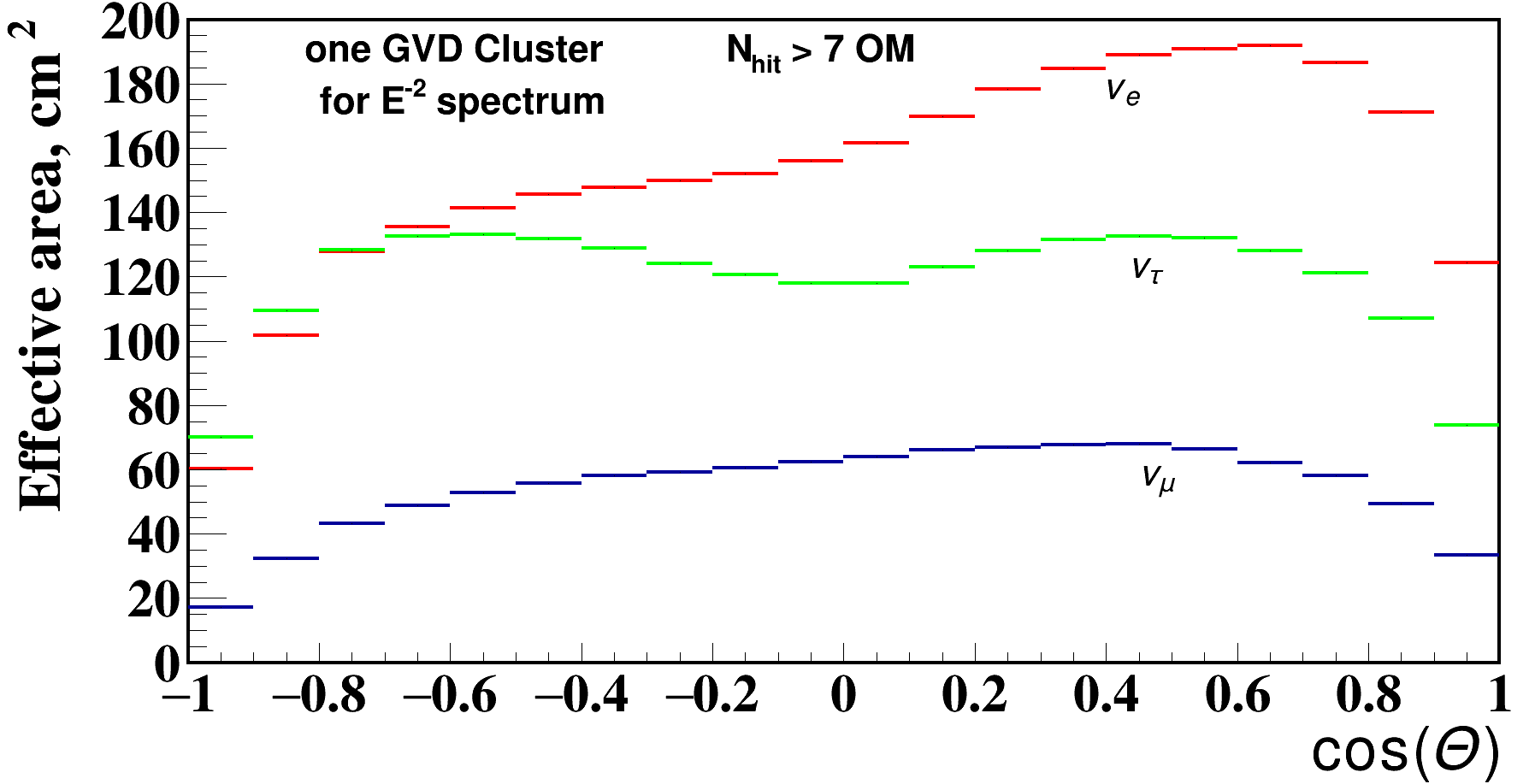}
    \includegraphics[width=.326\textwidth]{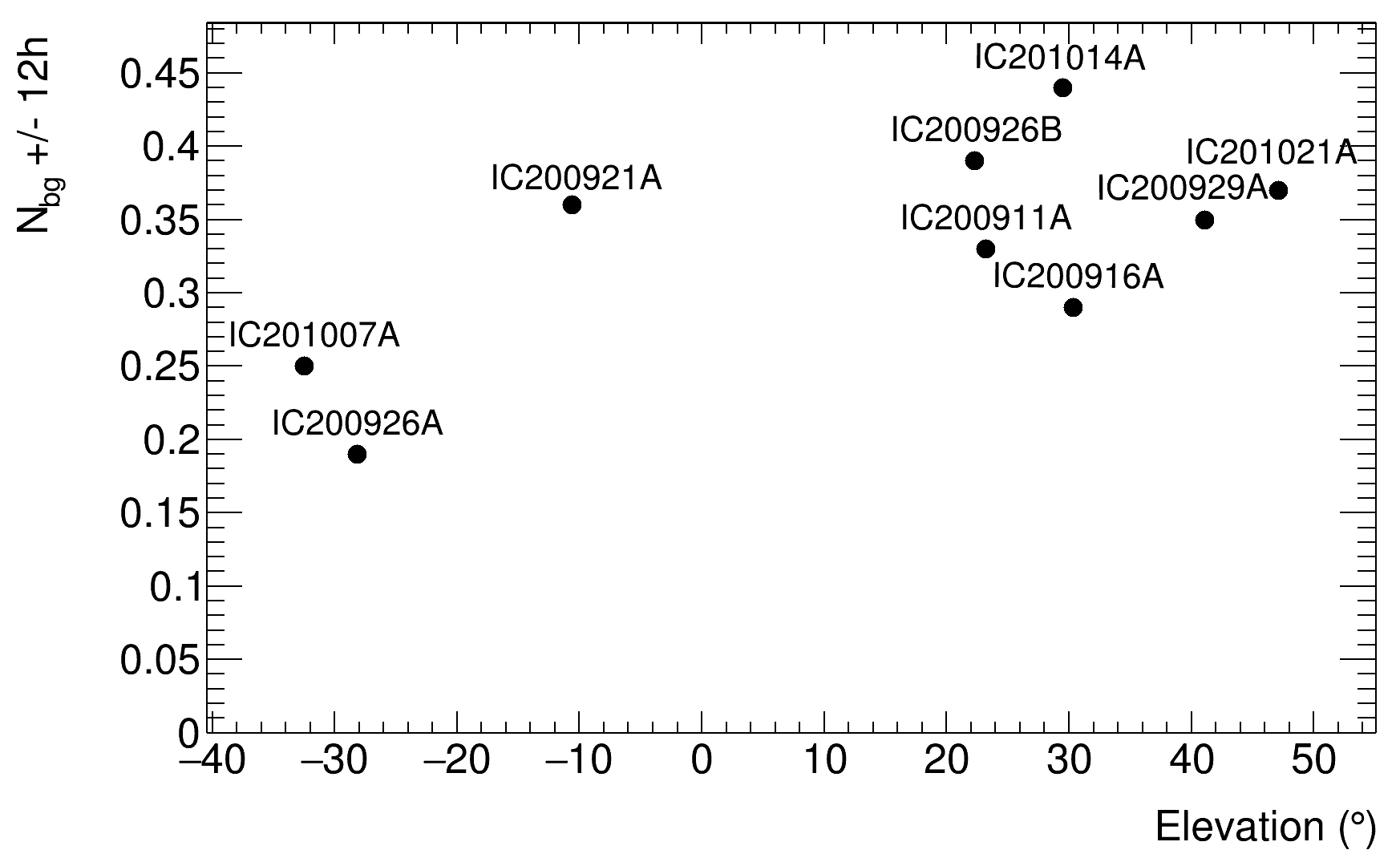}
	\includegraphics[width=.326\textwidth]{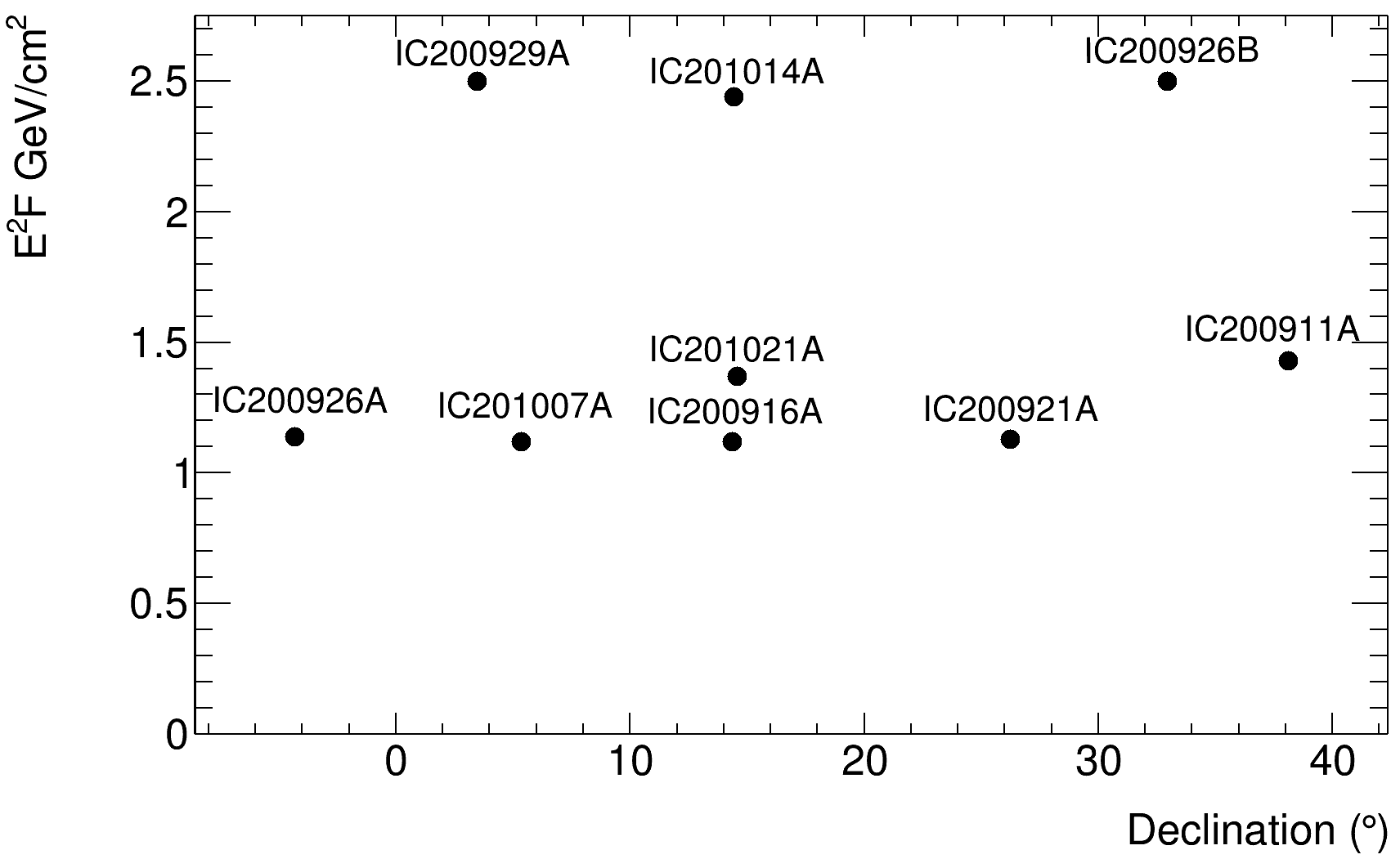}
	\caption{\label{fig:201120ICar} Left: Effective area for each neutrino flavor on cosine of zenith angle in the energy range 1 TeV--10 PeV for a single cluster in cascade mode reconstruction.
	Middle: Number of background events for nine fall alerts depending on source elevation i.e. (90-zenith angle). Right: Upper limits on fluence for these alerts depending on source declination. }
\end{figure}
In case of absence of a statistically significant excess of the number of coincidences above the expected background, upper limits on the number of expected signal events at the 90\% confidence level (C.L.) N$_{90\%}$ is calculated by the method of Feldman\,--\,Cousins statistics\cite{Feld}. Using N$_{90\%}$ and values of effective neutrino area \cite{casc_area1}, \cite{casc_area2} we compute upper limits at 90\% C.L. on the energy-dependent fluence for E$^{-2} $ spectrum assuming of an equal fraction of neutrino types in the total fluence and for given time window. The neutrino effective area for a single cluster is calculated as:
\begin{equation}  \label{eq:cone}
		A^{\nu}_{eff}(E_{\nu}, \theta_{\nu}, \phi_{\nu})= \frac {N_{sel}(E_{\nu}, \theta_{\nu}, \phi_{\nu})}{N_{gen}(E_{\nu}, \theta_{\nu}, \phi_{\nu})} \cdot V_{gen} \cdot (\rho N_{A}) \cdot \sigma(E_{\nu}) \cdot P_{Earth}(E_{\nu}, \theta_{\nu}),
\end{equation}
where $ N_{sel}(E_{\nu}, \theta_{\nu}, \phi_{\nu})$ and $N_{gen}(E_{\nu}, \theta_{\nu}, \phi_{\nu})$ are the number of selected and generated events in the generation volume of detector $ V_{gen} $, $\rho$ is the target density, $ N_A $ - Avogadro's number, $ \sigma(E_{\nu})$ - neutrino interaction cross section and $ P_{Earth} $ is the probability of neutrino to pass through the Earth without absorption. 
The dependence of the effective area for each neutrino flavor on cosine of zenith angle in the energy range 1 TeV--10 PeV is shown in Figure \ref{fig:201120ICar}, left (also see \cite{casc_area1}, \cite{casc_area2}) for a single cluster in cascade mode reconstruction. 

More details of the fluence calculations with the same data sets for the first nine IceCube alerts could be found in \cite{Astr_let}. We found zero cascades for all the alerts but three of them: IC200926B, IC200929A and IC201014A, for which one cascade was found in $\pm $12 h time interval for each alert but no events was found in the $\pm $1h time window. The probability of these events to belong to background is 0.32 (0.99 $\sigma$), 0.29 (1.11 $ \sigma $) and 0.36 (0.85 $ \sigma $), for each of the three alerts respectively. The complete Baikal-GVD data sample for 2019 (April 2019 -- February 2020) corresponds to 1495.15 live days of effective data accumulation by single cluster of the telescope and this sample was used to estimate the background.
Therefore, no statistically significant excess over the background was observed. (N$_{bg}$ values in $\pm$12 h see on Figure \ref{fig:201120ICar} (middle)). 
For mentioned three alerts the upper limits on fluence at 90\% C.L. are approximately 2.5 GeV/cm$^{2}$ in $\pm$12h, for other alerts this value is in the interval between 1--1.5 GeV/cm$^{2}$ in $\pm$12 h (Figure \ref{fig:201120ICar} (right)).  

In summary, the Baikal-GVD telescope data was analyzed for the first time in quasi-online neutrino alert tracking mode. No prompt correlations in direction and time were found with IceCube alerts between September 2020 to May 2021. The method of Feldman\,--\,Cousins statistics is applied to derive the upper limit on the number of events at the 90\% C.L. inside a cone 5$^{\circ}$ in time-window $\pm$ 12 h around the time of the alert.  Assuming an energy range between 1 TeV and 10 PeV with $E^{-2}$ spectrum and equal fluence in all flavors, the upper limits on the neutrino fluence is obtained using average effective area reported above. The results were obtained for cascade events using single clusters. This work was supported by the Ministry of Science and High Education of the Russian Federation within the financing program of large scientific projects of the Science National Project (grant no. 075-15- 2020-778).

\newpage

\section*{The Baikal-GVD Collaboration authors list}

\scriptsize
\noindent
{V.A.~Allakhverdyan}$^1$,
{A.D.~Avrorin}$^2$,
{A.V.~Avrorin}$^2$,
{V.M.~Aynutdinov}$^2$,
{R.~Bannasch}$^3$,
{Z.~Barda\v{c}ov\'{a}}$^4$,
{I.A.~Belolaptikov}$^1$,
{I.V.~Borina}$^1$,
{V.B.~Brudanin}$^{1\dagger}$,
{N.M.~Budnev}$^5$,
{V.Y.~Dik}$^1$,
{G.V.~Domogatsky}$^2$,
{A.A.~Doroshenko}$^2$,
{R.~Dvornick\'{y}}$^{1,4}$,
{A.N.~Dyachok}$^5$,
{Zh.-A.M.~Dzhilkibaev}$^2$,
{E.~Eckerov\'{a}}$^4$,
{T.V.~Elzhov}$^1$,
{L.~Fajt}$^6$,
{S.V.~Fialkovski}$^{7\dagger}$,
{A.R.~Gafarov}$^5$,
{K.V.~Golubkov}$^2$,
{N.S.~Gorshkov}$^1$,
{T.I.~Gress}$^5$,
{M.S.~Katulin}$^1$,
{K.G.~Kebkal}$^3$,
{O.G.~Kebkal}$^3$,
{E.V.~Khramov}$^1$,
{M.M.~Kolbin}$^1$,
{K.V.~Konischev}$^1$,
{K.A.~Kopa\'{n}ski}$^8$,
{A.V.~Korobchenko}$^1$,
{A.P.~Koshechkin}$^2$,
{V.A.~Kozhin}$^9$,
{M.V.~Kruglov}$^1$,
{M.K.~Kryukov}$^2$,
{V.F.~Kulepov}$^7$,
{Pa.~Malecki}$^8$,
{Y.M.~Malyshkin}$^1$,
{M.B.~Milenin}$^2$,
{R.R.~Mirgazov}$^5$,
{D.V.~Naumov}$^1$,
{V.~Nazari}$^1$,
{W.~Noga}$^8$,
{D.P.~Petukhov}$^2$,
{E.N.~Pliskovsky}$^1$,
{M.I.~Rozanov}$^{10}$,
{V.D.~Rushay}$^1$,
{E.V.~Ryabov}$^5$,
{G.B.~Safronov}$^2$,
{B.A.~Shaybonov}$^1$,
{M.D.~Shelepov}$^2$,
{F.~\v{S}imkovic}$^{1,4,6}$,
{A.E. Sirenko}$^1$,
{A.V.~Skurikhin}$^9$,
{A.G.~Solovjev}$^1$,
{M.N.~Sorokovikov}$^1$,
{I.~\v{S}tekl}$^6$,
{A.P.~Stromakov}$^2$,
{E.O.~Sushenok}$^1$,
{O.V.~Suvorova}$^2$,
{V.A.~Tabolenko}$^5$,
{B.A.~Tarashansky}$^5$,
{Y.V.~Yablokova}$^1$,
{S.A.~Yakovlev}$^3$
and
{D.N.~Zaborov}$^2$
\noindent

$^1$\textit{Joint Institute for Nuclear Research, Dubna, Russia}

$^2$\textit{Institute for Nuclear Research, Russian Academy of Sciences, Moscow, Russia}

$^3$\textit{EvoLogics GmbH, Berlin, Germany}

$^4$\textit{Comenius University, Bratislava, Slovakia}

$^5$\textit{Irkutsk State University, Irkutsk, Russia}

$^6$\textit{Czech Technical University in Prague, Prague, Czech Republic}

$^7$\textit{Nizhny Novgorod State Technical University, Nizhny Novgorod, Russia}

$^8$\textit{Institute of Nuclear Physics of Polish Academy of Sciences (IFJ~PAN), Krak\'{o}w, Poland}

$^9$\textit{Skobeltsyn Institute of Nuclear Physics, Moscow State University, Moscow, Russia}

$^{10}$\textit{St.~Petersburg State Marine Technical University, St.Petersburg, Russia}

\note[$\dagger$]{Deceased}

\end{document}